\documentclass[prd,nofootinbib
,reprint
]{revtex4-2}

\usepackage[utf8x]{inputenc}
\usepackage{enumitem}
\usepackage{hyperref}%
\usepackage{amsmath}
\usepackage{amsthm}
\usepackage{amssymb}
\usepackage{amsfonts}
\usepackage{mathrsfs}
\usepackage{bbm}
\usepackage{graphicx}
 \usepackage{dsfont}
\usepackage{braket}
\usepackage{tensor}
\usepackage{comment}
\usepackage{euscript}

\theoremstyle{definition}

\theoremstyle{definition}

\theoremstyle{definition}

\newcommand{\Osp}[2]{\mathrm{OSp}(#1|#2)} 
\newcommand{\UOsp}[2]{\mathrm{UOSp}(#1|#2)}

\newcommand{\C}{\mathbb{C}}

\newcommand{\calN}{\mathcal{N}}

\begin{document}
\title{Chiral loop quantum supergravity and black hole entropy}

\author{Konstantin Eder}
\email{konstantin.eder@gravity.fau.de}
\affiliation{Institute for Quantum Gravity (IQG), Department of Physics, 
Friedrich-Alexander-Universit\"at  Erlangen-N\"urnberg (FAU), Erlangen (Germany)}
\author{Hanno Sahlmann}
\email{hanno.sahlmann@gravity.fau.de}
\affiliation{Institute for Quantum Gravity (IQG), Department of Physics, 
Friedrich-Alexander-Universit\"at  Erlangen-N\"urnberg (FAU), Erlangen (Germany)}

%
%
\begin{abstract}
Recent work has shown that local supersymmetry on a spacetime boundary in $\mathcal{N}$-extended AdS supergravity in chiral variables implies coupling to a boundary $\Osp{\mathcal{N}}{2}_{\C}$ super Chern-Simons theory. 
We propose a way to calculate the entropy $S$ for the boundary, in the supersymmetric version of loop quantum gravity, for the minimal case $\mathcal{N}=1$. We calculate the dimensions of the quantum state spaces of $\UOsp{1}{2}$ super Chern-Simons theory with punctures, and analytically continue, for fixed quantum super area of the surface, to $\Osp{1}{2}_{\C}$. We find $S = a_H/4$ for large areas and determine the subleading correction.  



\end{abstract}

\maketitle


\section{Introduction}\label{superCartan geometry}
Since there are indications that horizons can meaningfully be assigned a thermodynamic entropy \cite{Bardeen:1973gs,Bekenstein:1973ur,Hawking:1975vcx}, the challenge is to explain it as the von Neumann-type entropy of a quantum description of the black hole. It has met with some measure of success in string theory (entropy of BPS black holes, for example \cite{Strominger:1996sh,Behrndt:1996jn}) and loop quantum gravity (entropy of isolated horizons, for example \cite{Smolin:1995vq,Rovelli:1996dv,Ashtekar:1997yu,Kaul:1998xv,Domagala:2004jt,Meissner:2004ju,Engle:2009vc,Agullo:2009eq}). Interestingly, the two approaches are very different in nature and the results concern disjoint families of black holes. 
The purpose of this letter is to report on recent work \cite{Eder:2022gge} that starts to bridge this gap. The theory considered is $\calN=1, D=4$ supergravity, on a manifold with a boundary with special boundary conditions. We calculate the entropy as the log of the size of the space of quantum states of a super Chern-Simons theory on the boundary, using analytic continuation from a compact real form, following \cite{Frodden:2012dq,Achour:2014eqa,Han:2014xna}. The Chern-Simons theory is constrained by the area of the boundary, and hence the entropy becomes area-dependent. 
We find that
\begin{equation}
\label{eq:1}
    S=\frac{a_H}{4l_p^2}+\mathcal{O}(\sqrt{a_H}/l_p)
\end{equation}
where $a_H$ is the diffeomorphism and gauge invariant measure of area\footnote{Bosonic area is not an observable since it is not gauge invariant. However, in a gauge in which the superelectric field has vanishing odd components, super area and bosonic area would agree.} in the supergeometric setting. 

We are using variables that were first proposed in \cite{Fulop:1993wi} and whose geometric meaning was recently clarified in \cite{Eder:2020erq,Eder:2021rgt,Eder:2022}. A super Chern-Simons theory as a source of entropy was first considered in \cite{Smolin:1995vq}. Supergravity with loop quantum gravity methods has also been considered in \cite{Ling:1999gn}, and using different variables in \cite{Bodendorfer:2011hs,Bodendorfer:2011pb,Bodendorfer:2011pc}. Our treatment follows \cite{Fulop:1993wi,Smolin:1995vq,Ling:1999gn,Gambini:1995db} in keeping supersymmetry manifest, but it goes further by making use of a detailed geometric analysis of the super Ashtekar connection and corresponding boundary conditions. It is also the first that is based on a detailed state counting in super Chern-Simons theory, as far as we know. Moreover, this is the first time that the Bekenstein-Hawking area law is derived and verified within the supersymmetric setting.

The quantum theory is obtained from a canonical formulation in terms of a supersymmetric generalization \cite{Fulop:1993wi,Eder:2020erq,Eder:2021rgt,Eder:2022} of the (chiral) Ashtekar connection $\mathcal{A}^+$ which can be obtained from a Holst modification of the McDowell Mansouri action \cite{Eder:2021rgt,Eder:2022}. The structure group in this formulation is $\Osp{1}{2}$.
We consider this theory in the presence of a causal boundary of spacetime playing the role of the horizon.The requirement of local supersymmetry also on the boundary \emph{uniquely} fixes a supersymmetric boundary term that is given by an $\Osp{1}{2}$ Chern-Simons theory and boundary conditions
\begin{equation}
    \underset{\raisebox{1pt}{$\Leftarrow$}}{F(\mathcal{A}^{+})}\propto \underset{\raisebox{1pt}{$\Leftarrow$}}{\mathcal{E}}
\end{equation}
linking curvature and super-electric field on the boundary. As in the non-supersymmetric case \cite{Ashtekar:1997yu}, the idea is to quantize bulk and boundary separately and couple them via the boundary condition. Field excitations in the bulk then couple to Chern-Simons defects in the boundary theory. Important elements of the bulk quantum theory (in fact, for $\mathcal{N}=1$ and $2$) 
including the definition of the graded holonomy-flux algebra, supersymmetric generalizations of spin networks, and the supersymmetric area operator can be defined \cite{Eder:2022gge}. $\Osp{1}{2}$ is non-compact, leading to various technical problems. We therefore start from the Chern-Simons theory of a compact real form of this group, $\UOsp{1}{2}$ and use analytical continuation in the corresponding Verlinde-type formula that is counting its states, following and generalizing \cite{Frodden:2012dq,Achour:2014eqa}.

 
We note that the present calculation is different from the string theory one in some respects. 
It applies to a large class of surfaces that carry local supersymmetry, whereas in string theory a more restrictive class of surfaces corresponding to BPS black holes is considered. Fermionic degrees of freedom
are taken directly into account in the present calculation, whereas they play no direct role in the string theory calculation. 


\section{The Holst-MacDowell-Mansouri action and chiral supergravity}\label{sec:Review}
We use the Cartan geometric description of pure AdS Holst-supergravity with $\mathcal{N}$-extended supersymmetry with $\mathcal{N}=1,2$. Details are in \cite{Eder:2022,Eder:2021rgt,Eder:2020erq} (see also \cite{Andrianopoli:2014aqa,Andrianopoli:2020zbl} using standard variables).\\
Pure AdS (Holst-)supergravity can be described in terms of a super Cartan geometry modeled on the super Klein geometry $(\mathrm{OSp}(\mathcal{N}|4),\mathrm{Spin}^+(1,3)\times\mathrm{SO}(\mathcal{N}))$ with super Cartan connection 
\begin{equation}
\mathcal{A}=e^IP_I+\frac{1}{2}\omega^{IJ}M_{IJ}+\frac{1}{2}\hat{A}_{rs}T^{rs}+\Psi_r^{\alpha}Q_{\alpha}^r
\label{D.eq:5.1.0}
\end{equation}
This connection can be used in order to formulate a Yang-Mills-type action principle for Holst-supergravity. One introduces a $\beta$-deformed inner product $\braket{\cdot\wedge\cdot}_{\beta}$ on $\mathfrak{g}\equiv\mathfrak{osp}(\mathcal{N}|4)$-valued differential forms on the underlying spacetime manifold $M$ with $\beta$ the Barbero-Immirzi parameter obtained via contraction of the $\mathrm{Ad}$-invariant supertrace with a $\beta$-dependent operator $\mathbf{P}_{\beta}$ on $\Omega^2(M,\mathfrak{g})$ (the precise form of this operator does not matter in what follows; for more details see \cite{Eder:2022,Eder:2021rgt}). Using this inner product, the Holst-MacDowell-Mansouri action of $\mathcal{N}$-extended pure Ads Holst-supergravity takes the form
\begin{equation}
S^{\beta}_{\text{H-MM}}(\mathcal{A})=\frac{L^2}{\kappa}\int_{M}{\braket{F(\mathcal{A})\wedge F(\mathcal{A})}_{\beta}}
\label{D.eq:6.4}
\end{equation}
with $F(\mathcal{A})$ the Cartan curvature of $\mathcal{A}$.

In the chiral limit of the theory corresponding to an imaginary $\beta=-i$, the action \eqref{D.eq:6.4} becomes manifestly invariant under an enlarged $\mathrm{Osp}(\mathcal{N}|2)_{\mathbb{C}}$-gauge symmetry: The operator $\mathbf{P}_{-i}$ decomposes as $\mathbf{P}_{-i}=\tilde{\mathbf{P}}_{-i}\circ\mathbf{P}^{\mathfrak{osp}(\mathcal{N}|2)}$ with $\mathbf{P}^{\mathfrak{osp}(\mathcal{N}|2)}:\,\mathfrak{osp}(\mathcal{N}|4)\rightarrow\mathfrak{osp}(\mathcal{N}|2)_{\mathbb{C}}$ the projection operator onto the (complexified) chiral sub superalgebra $\mathfrak{osp}(\mathcal{N}|2)_{\mathbb{C}}$ of $\mathfrak{g}$. Applying this projection operator on the super Cartan connection \eqref{D.eq:5.1.0} this yields the super Asthekar connection 
\begin{equation}
\mathcal{A}^{+}:=\mathbf{P}^{\mathfrak{osp}(\mathcal{N}|2)}\mathcal{A}=A^{+ i}T_i^{+}+\frac{1}{2}\hat{A}_{rs}T^{rs}+\psi^A_r Q_A^r
\label{D.eq:SuperAshtekarGeneral}
\end{equation}
In the chiral limit, the action becomes
\begin{align}
S^{\beta=-i}_{\text{H-MM}}(\mathcal{A})=\frac{i}{\kappa}\int_{M}{\left(\braket{F(\mathcal{A^+})\wedge\mathcal{E}}+\frac{1}{4L^2}\braket{\mathcal{E}\wedge\mathcal{E}}\right)}+S_{\text{bdy}}
\label{D.eq:7.8}
\end{align}
with $\mathcal{E}$ the super electric field canonically conjugate to the super Asthekar connection $\mathcal{A}^+$ and transforming under the Adjoint representation of $\mathrm{OSp}(\mathcal{N}|2)_{\mathbb{C}}$. $L$ is related to the cosmological constant, $L^2=-3/\Lambda_\text{cos}$. 
The boundary action is given by
\begin{equation}
S_{\text{bdy}}(\mathcal{A}^+)\!=\!\frac{k}{4\pi}\!\int_{H}{\braket{\mathcal{A}^+\wedge\mathrm{d}\mathcal{A}^+\!\!+\frac{1}{3}\mathcal{A}^+\!\wedge[\mathcal{A}^+\!\wedge\mathcal{A}^+]}}
\label{D.eq:7.12}
\end{equation}
i.e., the action of a $\mathrm{OSp}(\mathcal{N}|2)_{\mathbb{C}}$ super Chern-Simons theory with (complex) Chern-Simons level $k=i4\pi L^2/\kappa=-i12\pi/\kappa\Lambda_{\text{cos}}$. As discussed in detail in \cite{Eder:2022,Eder:2021rgt}, this boundary action arising from \eqref{D.eq:6.4} in the chiral limit is unique if one imposes supersymmetry invariance at the boundary (see also \cite{Andrianopoli:2014aqa,Andrianopoli:2020zbl}).

The decomposition of \eqref{D.eq:7.8} into a bulk and boundary action leads to an additional boundary condition coupling bulk and boundary degrees of freedom in order to ensure consistency with the equations of motion of the full theory,
\begin{equation}
    \underset{\raisebox{1pt}{$\Longleftarrow$}}{F(\mathcal{A}^{+})}=-\frac{1}{2L^2}\underset{\raisebox{1pt}{$\Leftarrow$}}{\mathcal{E}}
    \label{D.eq:4.11.2}
\end{equation}
where the arrow denotes the pullback of the respective fields to the boundary. 

\section{Quantum theory}
Since the phase space of AdS Holst-supergravity turns out to be a graded generalization of the purely bosonic theory, we quantize the theory adapting and generalizing tools from standard LQG. For more details, see \cite{Eder:2022}. 

\subsection{Super spin networks and the super area operator}\label{Section:chSUGRA-superspin}

In standard LQG, for the construction of the bulk Hilbert space $\mathfrak{H}^{\mathrm{bulk}}_{\gamma}$ associated to a graph $\gamma$ embedded in the spatial slices $\Sigma$ of the spacetime manifold $M=\mathbb{R}\times\Sigma$, one considers spin network states, a class of states invariant under local gauge transformations. They are constructed via contraction of matrix coefficients of irreducible representations of the underlying gauge group. 

For this it is crucial that the representations under consideration form a tensor category. Finite-dimensional irreducible representations of the orthosymplectic series $\mathrm{OSp}(\mathcal{N}|2)$ for $\mathcal{N}=1,2$ have been intensively studied (see e.g. \cite{Scheunert:1976wj,Scheunert:1976wi,Minnaert:1990sz,Berezin:1981}. For the case $\mathcal{N}=1$, these representations form a subcategory closed under the tensor product. The corresponding spin network states 
have been studied for instance in \cite{Gambini:1995db,Ling:1999gn}. 
For the case $\mathcal{N}=2$ the subclass of \emph{typical representations} form such a category (see \cite{Scheunert:1976wj}). 

We now describe the construction of the super spin network states for a suitable subclass $\EuScript{P}_{\mathrm{adm}}$ of irreducible representations (finite- or infinite-dimensional, possibly including those constructed in \cite{Eder:2022gge}) of $\mathrm{OSp}(\mathcal{N}|2)$ with $\mathcal{N}=1,2$. For any subset $\Vec{\pi}:=\{\pi_{e}\}_{e\in E(\gamma)}\subset\EuScript{P}_{\mathrm{adm}}$, we define the cylindrical function $T_{\gamma,\Vec{\pi},\Vec{m},\Vec{n}}$ via
\begin{equation}
    T_{\gamma,\Vec{\pi},\Vec{m},\Vec{n}}[\mathcal{A}^+]:=\prod_{e\in E(\gamma)}\tensor{\pi_e(h_e[\mathcal{A}^+])}{^{m_e}_{n_e}}
    \label{DH.eq:3.18}
\end{equation}
where, for any edge $e\in E(\gamma)$, $h_e[\mathcal{A}^+]$ denotes the super holonomy (parallel transport) of the connection $\mathcal{A}^+$ along $e$ (see \cite{Eder:2022,Eder:2021ans}) and $\tensor{(\pi_e)}{^{m_e}_{n_e}}$ denote certain matrix coefficients of the representation $\pi_e\in\EuScript{P}_{\mathrm{adm}}$. In order to get a gauge invariant state, at each vertex $v\in V(\gamma)$ of the graph $\gamma$, we have to contract \eqref{DH.eq:3.18} with an intertwiner $I_v$ projecting onto the trivial representation at any vertex. As a result, the so-constructed state transforms trivially under local gauge transformations and thus indeed forms a gauge-invariant state which we call a \emph{(gauge-invariant) super spin network state}. We take these states as a basis of the state space of the bulk theory. We assume that an inner product can be found that turns this space into a super Hilbert space $\mathfrak{H}^{\mathrm{bulk}}_{\gamma}$.
 
On the space of super spin networks, one can introduce a gauge-invariant operator in analogy to the area operator in ordinary LQG. More precisely, since the super electric field $\mathcal{E}$ defines a $\mathfrak{g}$-valued 2-form, for any oriented (semianalytic) surface $S$ embedded in $\Sigma$, one can define the \emph{graded} or \emph{super area} $\mathrm{gAr}(S)$ via 
\begin{equation}
\mathrm{gAr}(S):=\sqrt{2}\int_{S}{\sqrt{\braket{\mathcal{E}_S,\mathcal{E}_S}}}
\label{Chapter5:area1}
\end{equation}
where, in analogy to \cite{Eder:2018uzm,Corichi:2000dm,Ashtekar:2000hw}, $\mathcal{E}_S$ is defined as the unique $\mathfrak{g}$-valued function on $S$ such that $\iota^*_S\mathcal{E}=\mathcal{E}_S\,\mathrm{vol}_S$. For  the special case $\mathcal{N}=1$, the expression \eqref{Chapter5:area1} coincides with the super area as considered in \cite{Ling:1999gn}. Here, the prefactor $\sqrt{2}$ has been chosen such that in the case of vanishing fermionic degrees of freedom, the super area reduces to the standard area of $S$ in ordinary Riemannian geometry.

By definition, the quantity \eqref{Chapter5:area1} solely depends on the super electric field which defines a phase space variable. Thus, we can implement it in the quantum theory (see \cite{Eder:2022gge} for more details). As a result, for $\mathcal{N}=1$, it follows for instance in the case that the surface $S$ intersects the graph $\gamma$ of a (gauge-invariant) super spin network state $T_{\gamma,\Vec{\pi},\Vec{m},\Vec{n}}$ labeled by super spin quantum numbers $j\in\mathbb{C}$ corresponding to the principal series representations of $\mathrm{OSp}(1|2)$ as constructed in \cite{Eder:2022gge} in a single divalent vertex $v\in V(\gamma)$ that the action of super area operator is given by
\begin{equation}
\widehat{\mathrm{gAr}}(S) T_{\gamma,\Vec{\pi},\Vec{m},\Vec{n}}=-8\pi il_p^2\sqrt{j\left(j+\frac{1}{2}\right)}T_{\gamma,\Vec{\pi},\Vec{m},\Vec{n}}
\label{eq:3.3.49}
\end{equation}
with $j\in\mathbb{C}$ the superspin quantum number labeling the edge $e\in E(\gamma)$ intersecting the surface $S$. For $j\in\frac{N_0}{2}$, this coincides with the result of \cite{Ling:1999gn}.

According to \eqref{eq:3.3.49} the super area operator has complex eigenvalues which seems to be physically inconsistent. However, the principal series contains a subclass of irreducible representations with respect to which the super area operator becomes purely real. On the series of representations labeled by superspin quantum numbers of the form
\begin{align}
    j=-\frac{1}{4}+is\text{ with }s\in\mathbb{R}
    \label{eq:4.2.21}
\end{align}
the action of the super area operator takes the form
\begin{equation}
\widehat{\mathrm{gAr}}(S) T_{\gamma,\Vec{\pi},\Vec{m},\Vec{n}}=8\pi l_p^2\sqrt{s^2+\frac{1}{16}}\,T_{\gamma,\Vec{\pi},\Vec{m},\Vec{n}}
\label{eq:4.2.23}
\end{equation}
Super spin network states whose edges are labeled by $j$ satisfying \eqref{eq:4.2.21} are indeed eigenstates of the super area operator with real eigenvalues. Interestingly, this is in complete analogy to the bosonic theory \cite{Frodden:2012dq}.

\subsection{Coupling of boundary and bulk}\label{OutlookBHentropy}

As described in Section \ref{Section:chSUGRA-superspin} the quantum excitations of the bulk degrees of freedom are represented by super spin network states associated to the gauge supergroup $\mathrm{OSp}(\mathcal{N}|2)_{\mathbb{C}}$. On the other hand, in Section \ref{sec:Review}, we have explained that the boundary theory is described in terms of a $\mathrm{OSp}(\mathcal{N}|2)_{\mathbb{C}}$ super Chern-Simons theory. Hence, for a given finite graph $\gamma$ embedded in $\Sigma$, we define the Hilbert space $\mathfrak{H}_{\text{full},\gamma}$ w.r.t. $\gamma$ of the full theory as the tensor product
\begin{equation}
    \mathfrak{H}^{\text{full}}_{\gamma}=\mathfrak{H}^{\mathrm{bulk}}_{\gamma}\otimes\mathfrak{H}^{\text{bdy}}_{\gamma}
    \label{Gauss:fullTheory1}
\end{equation}
with $\mathfrak{H}^{\mathrm{bulk}}_{\gamma}$ the Hilbert space of the quantized bulk degrees of freedom as constructed in Section \ref{Section:chSUGRA-superspin} and $\mathfrak{H}^{\text{bdy}}_{\gamma}$ the Hilbert space corresponding to the quantized super Chern-Simons theory on the boundary.\\
On this Hilbert space, we have to implement the boundary condition \eqref{D.eq:4.11.2}. To this end, at each \emph{puncture} $p\in\EuScript{P}_{\gamma}:=\gamma\cap\Delta$, we choose a disk $D_{\epsilon}(p)$ on $\Delta$ around $p$ with radius $\epsilon>0$ and set 
\begin{equation}
    \mathcal{E}[\alpha](p):=\lim_{\epsilon\rightarrow 0}\int\limits_{D_{\epsilon}(p)}\braket{\alpha,\mathcal{E}},\quad F[\alpha](p):=\lim_{\epsilon\rightarrow 0}\int\limits_{D_{\epsilon}(p)}\braket{\alpha,F(\mathcal{A}^+)}
    \label{Gauss:fullTheory2}
\end{equation}
By definition, these quantities (or suitable functions thereof) can be promoted to well-defined operators in the quantum theory. Thus, \eqref{D.eq:4.11.2} yields the additional constraint equation    
\begin{equation}
\mathds{1}\otimes\widehat{F}_{\underline{A}}(p)=-\frac{2\pi i}{\kappa k}\widehat{\mathcal{E}}_{\underline{A}}(p)\otimes\mathds{1}   
\label{Gauss:fullTheory4}
\end{equation}
at each puncture $p\in\EuScript{P}_{\gamma}$, in analogy to the bosonic theory \cite{Engle:2009vc,Kaul:1998xv}. The quantized super electric flux $\widehat{\mathcal{E}}_{\underline{A}}(p)$ acts in terms of right- resp. left-invariant vector fields (see \cite{Eder:2022}). Hence, from \eqref{Gauss:fullTheory4}, we deduce that the Hilbert space of the quantized boundary degrees of freedom corresponds to the Hilbert space of a quantized super Chern-Simons theory on $\Delta$ with punctures $\EuScript{P}_{\gamma}$. This leads to the well-known (super)conformal blocks. 

\section{Entropy calculation}\label{sec:EntropyBH}

As discussed in Section \ref{OutlookBHentropy}, the boundary theory of chiral loop quantum supergravity for the case $\mathcal{N}=1$ is described by a quantized super Chern-Simons theory with punctures and gauge supergroup $\mathrm{OSp}(1|2)_{\mathbb{C}}$ as well as complex Chern-Simons level. Hence, to the boundary one can associate an entropy in terms of the number of Chern-Simons degrees of freedom generated by the super spin network edges piercing the boundary. Unfortunately, the (super) Chern-Simons theory with complex and non-compact gauge group is not well-known. Moreover, it is not clear how to deal with the fact that the Chern-Simons level is purely imaginary. Interestingly, similar issues also seem to arise in the context of boundary theories in string theory \cite{Mikhaylov:2014aoa}.
We therefore adapt the strategy of \cite{Achour:2014eqa} in the context of the purely bosonic theory by studying a specific compact real form of $\mathrm{OSp}(1|2)_{\mathbb{C}}$ and then performing an analytic continuation to the corresponding complex Lie supergroup. 

\subsection{Super characters of $\mathrm{UOSp(1|2)}$ and the Verlinde formula}
Let us consider the Chern-Simons theory with compact gauge supergroup given by the unitary orthosymplectic group $\mathrm{UOSp}(1|2)=\mathrm{U}(1|2)\cap\mathrm{OSp}(1|2)$ and integer Chern-Simons level $k=-12\pi/\kappa\Lambda_{\mathrm{cos}}$ and punctures labeled by finite-dimensional irreducible representations $\Vec{j}$ of $\mathrm{UOSp}(1|2)$ with $j\in\frac{\mathbb{N}_0}{2}$. We compute the number $\mathcal{N}_k(\Vec{j})$ of Chern-Simons degrees of freedom given by the dimension of the superconformal blocks. 
To simplify the discussion, we assume that the boundary $H$ is topologically of the form $\mathbb{R}\times\mathbb{S}^2$, that is, the $2$-dimensional slices $\Delta_t$ are topologically equivalent to $2$-spheres. Furthermore, we consider the limit $k\rightarrow\infty$ corresponding to a vanishing cosmological constant $\Lambda_{\mathrm{cos}}$. Under these assumptions, the number of microstates $\mathcal{N}_{\infty}(\Vec{j})$  is given by the number of $\mathrm{UOSp}(1|2)$ gauge-invariant states, i.e., it can be identified with the number of trivial subrepresentations contained in the tensor product representation $\bigotimes_j \pi_j$. In this way, by subdividing $\Vec{j}$ into $p\leq n$ subfamilies $(n_l,j_l)$, $l=1,\ldots,p$, consisting of $0<n_l\leq n$ punctures labeled by $j_l\in\Vec{j}$, one finds that the dimension of the conformal block in the limit $k\rightarrow\infty$ can be computed via the following integral formula
\begin{align}
    \mathcal{N}_{\infty}(\{n_l,j_l\})=&\frac{1}{2\pi}\int_0^{\pi}\mathrm{d}\theta\,\sin^2(2\theta)\prod_{l=1}^p\left(\frac{\cos(d_{j_l}\theta)}{\cos\theta}\right)^{n_l}\nonumber\times\\
    &\times\left[4-n+\sum_{i=1}^p n_i d_{j_l}\frac{\tan(d_{j_i}\theta)}{\tan\theta}\right]
    \label{eq:VerlindeFormula}
\end{align}
with $d_j:=4j+1$ the dimension of the spin-$j$ representation of $\mathrm{UOSp}(1|2)$. 

\subsection{The monochromatic case}
We use \eqref{eq:VerlindeFormula} to compute the entropy associated to the boundary by performing an analytic continuation $\mathrm{OSp}(1|2)_{\mathbb{C}}$ of chiral LQSG: We replace the superspin quantum numbers $j\in\Vec{j}$ in \eqref{eq:VerlindeFormula} by $j\rightarrow -\frac{1}{4}+is$, i.e., quantum numbers corresponding to the principal series with respect to which the super area operator has purely real eigenvalues. 
Consider the monochromatic case, i.e., assume that the punctures on the boundary are all labeled by the same super spin quantum number $j$. Then, by replacing $j\rightarrow-\frac{1}{4}+is$ for some $s\in\mathbb{R}_{>0}$ in \eqref{eq:VerlindeFormula} for the special case $p=1$ and using $d_j=i4s=:i\tilde{s}$ as well as $\cos(ix)=\cosh(x)$ and $\sin(ix)=i\sinh(x)$, one finds that an analytically continued version of \eqref{eq:VerlindeFormula} is given by the following contour integral
\begin{align}
  \mathcal{I}_{\infty}=\frac{1}{2\pi}\int_{\mathcal{C}}\mathrm{d}z&\,\mu(z)\left(4-n\left[1+\tilde{s}\frac{\tan(\tilde{s}z)}{\tanh z}\right]\right)\times\nonumber\\
    &\times\exp\left(n\ln\left(\frac{\cos(\tilde{s} z)}{\cosh z}\right)\right)
    \label{eq:5.2.1}
\end{align}
with density $\mu(z):=i\sinh^2(2z)$. Here, $\mathcal{C}$ denotes a contour from 0 to $i\pi$. 
We evaluate the integral formula \eqref{eq:5.2.1} in the macroscopic limit corresponding to the limit $s\rightarrow\infty$ and $n\rightarrow\infty$, i.e., large color and large number of punctures. In this limit, one can then apply the method of steepest descent, so we arrange $\mathcal{C}$ to go through all the (non-degenerate) critical points $z_c$, in the direction of steepest descent, of the ``action''
\begin{equation}
    \mathcal{S}(z)=\ln\left(\frac{\cos(\tilde{s} z)}{\cosh z}\right)
    \label{eq:5.2.2}
\end{equation}
located along the imaginary axis and lying between 0 and $i\pi$. $\mathcal{S}'(z_c)=0$ gives
\begin{equation}
    \tilde{s}\tan(\tilde{s}z_c)=-\tanh(z_c)
    \label{eq:5.2.4}
\end{equation}
For critical points lying on the imaginary axis, in the macroscopic limit an approximate solution to Eq. \eqref{eq:5.2.4} is given by $z_c=i(\frac{\pi}{2}-\epsilon)$ for some small $\epsilon$ of order $\epsilon=o(\tilde{s}^{-1})$. Inserting this into the action \eqref{eq:5.2.2}, one finds
\begin{equation}
    \mathcal{I}_{\infty}=\sqrt{\frac{2}{\pi}}\frac{1}{32s^3\sqrt{n}}\left(\frac{2s}{e}\right)^n\exp\left(\frac{a_H}{4}-i\frac{\pi}{2}\right)
    \label{eq:5.2.10}
\end{equation}
with $a_H=8\pi ns$ the super area of the boundary in the monochromatic case (see Eq. \eqref{eq:4.2.23}). The analytic continuation of the state sum acquires an additional complex phase which seems counter intuitive. This is, however, in analogy to the bosonic theory \cite{Achour:2014eqa} where it is argued that one instead needs to consider the modulus of \eqref{eq:5.2.10}. Doing so, for indistinguishable punctures the entropy $S:=\ln(|\mathcal{I}_{\infty}|/n!)$ of the boundary is given by
\begin{equation}
    S=\frac{a_H}{4l_p^2}+\nu\ln\left(\frac{2\sigma}{\nu}\right)\frac{\sqrt{a_H}}{l_p}-2\ln\left(\frac{a_H}{l_p^2}\right)+\mathcal{O}(1)
    \label{eq:5.2.14}
\end{equation}
where $\nu,\sigma>0$ are some numerical coefficients such that in the macroscopic limit $n=\nu\frac{\sqrt{a_H}}{l_p}$ as well as $s=\sigma\frac{\sqrt{a_H}}{l_p}$. The highest order term in \eqref{eq:5.2.14} exactly reproduces the Bekenstein-Hawking area law. This is a very intriguing result and follows here directly from the analytically continued Verlinde formula \eqref{eq:5.2.1}. In particular, we did not have to make any choices beyond working with the chiral theory and no parameter needed to be adjusted by hand. Moreover, this confirms the results of \cite{Achour:2014eqa} in the bosonic theory and supports the hypothesis that in the context of complex variables the entropy can be derived via an analytic continuation starting from a compact real form of the complex gauge group.

\subsection{The multi-color case}
Let us finally very briefly summarize the case of $j$ varying over the punctures. The analytical continuation of \eqref{eq:VerlindeFormula} is then given by 
\begin{align}
  \mathcal{I}_{\infty}=\frac{1}{2\pi}\int_{\mathcal{C}}\mathrm{d}z&\,\mu(z)\exp\left(\sum_{l=1}^p n_l\ln\left(\frac{\cos(\tilde{s}_l z)}{\cosh z}\right)\right)\times\nonumber\\
    &\times\left[4-n-\sum_{i=1}^p n_i\tilde{s}_i\frac{\tan(\tilde{s}_i z)}{\tanh z}\right]
    \label{eq:5.3.1}
\end{align}
Again, in the macroscopic limit, we can evaluate \eqref{eq:5.3.1} by using the method of steepest decent. In this way, following the same steps as in the previous section, one finds that the entropy, at highest order, associated to the boundary is given by (for more details see \cite{Eder:2022gge})
\begin{equation}
    S=\ln(|\mathcal{I}_{\infty}|/n!)=\frac{a_H}{4l_p^2}+\ldots
    \label{eq:5.3.8}
\end{equation}
and thus indeed corresponds to the Bekenstein-Hawking area law. The lower order quantum corrections can be computed similarly to the monochromatic case.


 \section{Discussion and outlook}

There are several surprises that came together to yield the Bekenstein-Hawking area law:
Boundary theory and boundary conditions are uniquely fixed from the requirement of supersymmetry; the boundary theory is a Chern-Simons theory and it couples to the bulk just as for isolated horizons in the non-supersymmetric theory; there is a compact real form of $\Osp{1}{2}_{\mathbb{C}}$ that one can find a Verlinde-type formula for; that $\Osp{1}{2}_{\mathbb{C}}$ possesses representations with the right properties to carry out the analytic continuation and that the Verlinde formula allows it. Also, the Chern-Simons level is proportional to the inverse of the cosmological constant. Therefore the large-$k$ limit makes physical sense and one does not have to deal with the intricacies of \emph{quantum deformations} of super groups. Finally, the only change in comparison to the non-supersymmetric theory in highest order turns out to be a factor of 2 in the exponent which can be easily incorporated into the picture by using the area eigenvalue of two-sided punctures at the horizon.    


There are several places where our arguments are not as stringent as they should be, and there are some open questions: The bulk quantum theory is not complete, and the quantum theory for the boundary Chern-Simons theory for the non-compact supergroup and at imaginary level is not known directly.   Moreover, we have not based our consideration on a theory of isolated horizons, since it has not been worked out yet for supergravity theories, as far as we know. Finally, the right-handed supersymmetry constraint has not been implemented in a direct way. Rather, our assumption is that it does not significantly reduce the number of surface states, as is assumed for the Hamilton constraint in the bosonic theory.  

In addition, it would be interesting to extend the theory to extended supersymmetry, $\calN>1$. We have a reasonably good understanding of how the bulk theory would look like for $\calN=2$. Based on this it seems feasible to extend the entropy calculation to more physically realistic models with $\calN=2$. Complementary to this, it would be very desirable to complete a calculation for BPS black holes as considered in string theory \cite{Strominger:1996sh,Behrndt:1996jn}.

Finally, it is very interesting to note that $\Osp{m}{n}_\mathbb{C}$ super Chern-Simons theories at complex level $k$ show up as boundary theories in string theory \cite{Mikhaylov:2014aoa}.  It would be great to better understand the possible connections to the present work in general, and in particular between the analytic continuation considered there and the one we used.   


\begin{acknowledgements}
We would like to thank Lee Smolin for communications at an early stage of this work and in particular for his interest in application of loop quantum gravity methods to supersymmetric black holes which was part of the motivation of this work. 
K.E. thanks the German Academic Scholarship Foundation (Studienstiftung des Deutschen Volkes) for financial support. H.S. acknowledges the contribution of the COST Action CA18108.
\end{acknowledgements}



\begin{thebibliography}{56}
\addcontentsline{toc}{chapter}{Bibliography}

\bibitem{Eder:2022gge}
K.~Eder and H.~Sahlmann,
``Toward black hole entropy in chiral loop quantum supergravity,''
Phys. Rev. D \textbf{106} (2022) no.2, 026001
doi:10.1103/PhysRevD.106.026001
[arXiv:2204.01661 [gr-qc]].

\bibitem{Bardeen:1973gs}
J.~M.~Bardeen, B.~Carter and S.~W.~Hawking,
``The Four laws of black hole mechanics,''
Commun. Math. Phys. \textbf{31} (1973), 161-170
doi:10.1007/BF01645742

\bibitem{Bekenstein:1973ur}
J.~D.~Bekenstein,
``Black holes and entropy,''
Phys. Rev. D \textbf{7} (1973), 2333-2346
doi:10.1103/PhysRevD.7.2333

\bibitem{Hawking:1975vcx}
S.~W.~Hawking,
``Particle Creation by Black Holes,''
Commun. Math. Phys. \textbf{43} (1975), 199-220
[erratum: Commun. Math. Phys. \textbf{46} (1976), 206]
doi:10.1007/BF02345020

\bibitem{Strominger:1996sh}
A.~Strominger and C.~Vafa,
``Microscopic origin of the Bekenstein-Hawking entropy,''
Phys. Lett. B \textbf{379} (1996), 99-104
doi:10.1016/0370-2693(96)00345-0
[arXiv:hep-th/9601029 [hep-th]].

\bibitem{Behrndt:1996jn}
K.~Behrndt, G.~Lopes Cardoso, B.~de Wit, R.~Kallosh, D.~Lust and T.~Mohaupt,
``Classical and quantum N=2 supersymmetric black holes,''
Nucl. Phys. B \textbf{488} (1997), 236-260
doi:10.1016/S0550-3213(97)00028-X
[arXiv:hep-th/9610105 [hep-th]].


\bibitem{Smolin:1995vq}
L.~Smolin,
``Linking topological quantum field theory and nonperturbative quantum gravity,''
J. Math. Phys. \textbf{36}, 6417-6455 (1995)
doi:10.1063/1.531251
[arXiv:gr-qc/9505028 [gr-qc]].

\bibitem{Rovelli:1996dv}
C.~Rovelli,
``Black hole entropy from loop quantum gravity,''
Phys. Rev. Lett. \textbf{77}, 3288-3291 (1996)
doi:10.1103/PhysRevLett.77.3288
[arXiv:gr-qc/9603063 [gr-qc]].

\bibitem{Ashtekar:1997yu}
A.~Ashtekar, J.~Baez, A.~Corichi and K.~Krasnov,
``Quantum geometry and black hole entropy,''
Phys. Rev. Lett. \textbf{80}, 904-907 (1998)
doi:10.1103/PhysRevLett.80.904
[arXiv:gr-qc/9710007 [gr-qc]].


\bibitem{Kaul:1998xv}
R.~K.~Kaul and P.~Majumdar,
``Quantum black hole entropy,''
Phys. Lett. B \textbf{439} (1998), 267-270
doi:10.1016/S0370-2693(98)01030-2
[arXiv:gr-qc/9801080 [gr-qc]].

\bibitem{Domagala:2004jt}
M.~Domagala and J.~Lewandowski,
``Black hole entropy from quantum geometry,''
Class. Quant. Grav. \textbf{21}, 5233-5244 (2004)
doi:10.1088/0264-9381/21/22/014
[arXiv:gr-qc/0407051 [gr-qc]].

\bibitem{Meissner:2004ju}
K.~A.~Meissner,
``Black hole entropy in loop quantum gravity,''
Class. Quant. Grav. \textbf{21}, 5245-5252 (2004)
doi:10.1088/0264-9381/21/22/015
[arXiv:gr-qc/0407052 [gr-qc]].


\bibitem{Engle:2009vc}
J.~Engle, A.~Perez and K.~Noui,
``Black hole entropy and SU(2) Chern-Simons theory,''
Phys. Rev. Lett. \textbf{105}, 031302 (2010)
doi:10.1103/PhysRevLett.105.031302
[arXiv:0905.3168 [gr-qc]].

\bibitem{Agullo:2009eq}
I.~Agullo, G.~J.~Fernando Barbero, E.~F.~Borja, J.~Diaz-Polo and E.~J.~S.~Villasenor,
``The Combinatorics of the SU(2) black hole entropy in loop quantum gravity,''
Phys. Rev. D \textbf{80} (2009), 084006
doi:10.1103/PhysRevD.80.084006
[arXiv:0906.4529 [gr-qc]].

\bibitem{Fulop:1993wi}
G.~Fulop,
``About a superAshtekar-Renteln ansatz,''
Class. Quant. Grav. \textbf{11}, 1-10 (1994)
doi:10.1088/0264-9381/11/1/005
[arXiv:gr-qc/9305001 [gr-qc]].

\bibitem{Eder:2022}
K.~Eder,
``Super Cartan geometry and loop quantum supergravity,''
FAU Forschungen, Reihe B, Medizin, Naturwissenschaft, Technik Band 40 (2022). Erlangen: FAU University Press. doi:10.25593/978-3-96147-530-8.



\bibitem{Eder:2021rgt}
K.~Eder and H.~Sahlmann,
``Holst-MacDowell-Mansouri action for (extended) supergravity with boundaries and super Chern-Simons theory,''
JHEP \textbf{07} (2021), 071
doi:10.1007/JHEP07(2021)071
[arXiv:2104.02011 [gr-qc]].


\bibitem{Eder:2020erq}
K.~Eder,
``Super Cartan geometry and the super Ashtekar connection,''
[arXiv:2010.09630 [gr-qc]].

\bibitem{Bodendorfer:2011hs}
N.~Bodendorfer, T.~Thiemann and A.~Thurn,
``Towards Loop Quantum Supergravity (LQSG),''
Phys. Lett. B \textbf{711}, 205-211 (2012)
doi:10.1016/j.physletb.2012.04.003
[arXiv:1106.1103 [gr-qc]].

\bibitem{Bodendorfer:2011pb}
N.~Bodendorfer, T.~Thiemann and A.~Thurn,
``Towards Loop Quantum Supergravity (LQSG) I. Rarita-Schwinger Sector,''
Class. Quant. Grav. \textbf{30}, 045006 (2013)
doi:10.1088/0264-9381/30/4/045006
[arXiv:1105.3709 [gr-qc]].

\bibitem{Bodendorfer:2011pc}
N.~Bodendorfer, T.~Thiemann and A.~Thurn,
``Towards Loop Quantum Supergravity (LQSG) II. p-Form Sector,''
Class. Quant. Grav. \textbf{30}, 045007 (2013)
doi:10.1088/0264-9381/30/4/045007
[arXiv:1105.3710 [gr-qc]].



\bibitem{Andrianopoli:2014aqa}
L.~Andrianopoli and R.~D'Auria,
``N=1 and N=2 pure supergravities on a manifold with boundary,''
JHEP \textbf{08} (2014), 012
doi:10.1007/JHEP08(2014)012
[arXiv:1405.2010 [hep-th]].

\bibitem{Andrianopoli:2020zbl}
L.~Andrianopoli, B.~L.~Cerchiai, R.~Matrecano, O.~Miskovic, R.~Noris, R.~Olea, L.~Ravera and M.~Trigiante,
``$ \mathcal{N} $ = 2 AdS$_{4}$ supergravity, holography and Ward identities,''
JHEP \textbf{02} (2021), 141
doi:10.1007/JHEP02(2021)141
[arXiv:2010.02119 [hep-th]].


\bibitem{Frodden:2012dq}
E.~Frodden, M.~Geiller, K.~Noui and A.~Perez,
``Black Hole Entropy from complex Ashtekar variables,''
EPL \textbf{107} (2014) no.1, 10005
doi:10.1209/0295-5075/107/10005
[arXiv:1212.4060 [gr-qc]].

\bibitem{Han:2014xna}
M.~Han,
``Black Hole Entropy in Loop Quantum Gravity, Analytic Continuation, and Dual Holography,''
[arXiv:1402.2084 [gr-qc]].

\bibitem{Achour:2014eqa}
J.~Ben Achour, A.~Mouchet and K.~Noui,
``Analytic Continuation of Black Hole Entropy in Loop Quantum Gravity,''
JHEP \textbf{06} (2015), 145
doi:10.1007/JHEP06(2015)145
[arXiv:1406.6021 [gr-qc]].


\bibitem{Berezin:1981}
F.~A.~Berezin, V.~N.~Tolstoy,
``The group with Grassmann structure UOSP(1|2),''
Commun. Math. Phys. 78 (1981) 409.

\bibitem{Scheunert:1976wj}
M.~Scheunert, W.~Nahm and V.~Rittenberg,
``Irreducible Representations of the OSP(2,1) and SPL(2,1) Graded Lie Algebras,''
J. Math. Phys. \textbf{18} (1977), 155
doi:10.1063/1.523149

\bibitem{Scheunert:1976wi}
M.~Scheunert, W.~Nahm and V.~Rittenberg,
``Graded Lie Algebras: Generalization of Hermitian Representations,''
J. Math. Phys. \textbf{18} (1977), 146
doi:10.1063/1.523148

\bibitem{Minnaert:1990sz}
P.~Minnaert and M.~Mozrzymas,
``Racah-Wigner calculus for the superrotation algebra. 1.,''
J. Math. Phys. \textbf{33} (1992), 1582-1593
doi:10.1063/1.529683

\bibitem{Gambini:1995db}
R.~Gambini, O.~Obregon and J.~Pullin,
``Towards a loop representation for quantum canonical supergravity,''
Nucl. Phys. B \textbf{460} (1996), 615-631
doi:10.1016/0550-3213(95)00582-X
[arXiv:hep-th/9508036 [hep-th]].

\bibitem{Ling:1999gn}
Y.~Ling and L.~Smolin,
``Supersymmetric spin networks and quantum supergravity,''
Phys. Rev. D \textbf{61} (2000), 044008
doi:10.1103/PhysRevD.61.044008
[arXiv:hep-th/9904016 [hep-th]].

\bibitem{Eder:2021ans}
K.~Eder,
``Super fiber bundles, connection forms, and parallel transport,''
J. Math. Phys. \textbf{62} (2021) no.6, 063506
doi:10.1063/5.0044343
[arXiv:2101.00924 [math.DG]].

\bibitem{Eder:2018uzm}
K.~Eder and H.~Sahlmann,
``Quantum theory of charged isolated horizons,''
Phys. Rev. D \textbf{97} (2018) no.8, 086016
doi:10.1103/PhysRevD.97.086016
[arXiv:1801.00747 [gr-qc]].

\bibitem{Corichi:2000dm}
A.~Corichi, U.~Nucamendi and D.~Sudarsky,
``Einstein-Yang-Mills isolated horizons: Phase space, mechanics, hair and conjectures,''
Phys. Rev. D \textbf{62} (2000), 044046
doi:10.1103/PhysRevD.62.044046
[arXiv:gr-qc/0002078 [gr-qc]].

\bibitem{Ashtekar:2000hw}
A.~Ashtekar, S.~Fairhurst and B.~Krishnan,
``Isolated horizons: Hamiltonian evolution and the first law,''
Phys. Rev. D \textbf{62} (2000), 104025
doi:10.1103/Phys
RevD.62.104025
[arXiv:gr-qc/0005083 [gr-qc]].

\bibitem{Mikhaylov:2014aoa}
V.~Mikhaylov and E.~Witten,
``Branes And Supergroups,''
Commun. Math. Phys. \textbf{340} (2015) no.2, 699-832
doi:10.1007/s00220-015-2449-y
[arXiv:1410.1175 [hep-th]].



\end{thebibliography}
\end{document}